\documentclass[aps,prc,twocolumn,superscriptaddress,nofootinbib,longbibliography,floatfix,10pt]{revtex4-2}

\usepackage{graphicx}
\usepackage{dcolumn}
\usepackage{bm}
\usepackage{morefloats}
\usepackage{multirow}
\usepackage{amssymb}
\usepackage{amsmath}
\usepackage{xcolor}
\usepackage{longtable}
\usepackage{fix-cm}
\usepackage{verbatim}
\usepackage{mathptmx} % rm & math
\usepackage[T1]{fontenc}
\usepackage[colorlinks,allcolors=blue]{hyperref}

\makeatletter
\def\NAT@def@citea{\def\@citea{\NAT@separator}}
\makeatother

\begin{document}

\title{Description of multinucleon transfer mechanism for ${}^{48} \mathrm{Ca}+{ }^{244} \mathrm{Pu}$ and ${ }^{86} \mathrm{Kr}+{}^{198} \mathrm{Pt}$ reactions in quantal transport approach}

\author{M. Arik}
\affiliation{Physics Department, Middle East Technical University, 06800 Ankara, Turkey}
\author{S. Ayik}\email{ayik@tntech.edu}
\affiliation{Physics Department, Tennessee Technological University, Cookeville, TN 38505, USA}
\author{O. Yilmaz}
\affiliation{Physics Department, Middle East Technical University, 06800 Ankara, Turkey}
\author{A. S. Umar}
\affiliation{Department of Physics and Astronomy, Vanderbilt University, Nashville, TN 37235, USA}

\date{\today}

\begin{abstract}
% the extra lines centers the abstract like the final PRC document
\edef\oldrightskip{\the\rightskip}
\begin{description}
\rightskip\oldrightskip\relax
\setlength{\parskip}{0pt} % no skip between items
\item[Background] Multinucleon transfer (MNT) reactions involving heavy
projectile and target combinations stand as a promising method for synthesizing
new neutron-rich exotic nuclei which may not be possible using hot or cold
fusion reactions or fragmentation. Exploring the mechanisms behind MNT reactions
is essential and it requires a comprehensive theoretical
framework that can explain the physical observables in these reactions.
\item[Purpose] This work aims to show that the quantal diffusion approach
based on the stochastic mean field (SMF) theory is capable of explaining the reaction dynamics
observed in MNT reactions. Primary product mass distributions
in $^{48}$Ca+$^{244}$Pu reaction at E$_\text{{c.m.}}= $ 203.2 MeV and
$^{86}$Kr+$^{198}$Pt reaction at E$_\text{{c.m.}} = $ 324.2 MeV are calculated
and compared with the available experimental data.
\item[Methods] In
this work, we utilize the time-dependent Hartree-Fock (TDHF) calculations to
analyze the mean-field reaction dynamics computationally in the reactions
${}^{48}\mathrm{Ca}+{}^{244}\mathrm{Pu}$ and
${}^{86}\mathrm{Kr}+{}^{198}\mathrm{Pt}$ for a broad range of initial angular
momentum. Quantal transport description based on the SMF approach is used to
calculate quantal diffusion coefficients and mass variances in
${}^{48}\mathrm{Ca}+{}^{244}\mathrm{Pu}$ and
${}^{86}\mathrm{Kr}+{}^{198}\mathrm{Pt}$ systems. The primary products arising
from quasi-fission reactions are described by joint probability distribution in
the SMF approach and those arising from fusion-fission are estimated by using
the statistical deexcitation code GEMINI++.
\item[Results] Mean
values of charge and mass numbers, scattering angles of the primary reaction
products, and the total kinetic energies after the collision are calculated within the
TDHF framework for a broad range of initial angular momenta. Throughout all the
collisions, drift toward the mass symmetry and large mass dispersion associated
with this drift are observed. The calculated primary fragment and mass
distributions using the SMF approach successfully explain experimental
observations for the ${}^{48}\mathrm{Ca}+{}^{244}\mathrm{Pu}$ and
${}^{86}\mathrm{Kr}+{}^{198}\mathrm{Pt}$ systems.
\item[Conclusions]
The primary mass distributions, mean values of binary products, and mass
dispersions are determined and results are compared with the available
experimental data. The observed agreement between the experimental data and SMF
results highlights the effectiveness of the quantal diffusion mechanism based on
the SMF approach, which does not include any adjustable parameters other than
standard parameters of Skyrme energy density functional.
\end{description}
\end{abstract}

\maketitle

\section{Introduction}\label{sec_1}
Over the last decade, the number of isotopes in the chart of nuclides has
increased significantly. Although there has been a great effort to synthesize
superheavy elements and exotic isotopes such as neutron-rich or proton-rich
isotopes, the relatively low cross sections of these isotopes still impose a
difficult challenge in heavy ion experiments. Superheavy elements produced so
far have been a product of either hot~\cite{oganessian2006,itkis2022} or cold
fusion~\cite{hofmann2000} reactions, wherein the highly excited compound, nuclei
de-excite mostly by neutron emission and secondary fission. As an alternative,
multinucleon transfer (MNT) in heavy-ion collisions is a promising approach for
the synthesis of new neutron-rich nuclei. For this purpose, MNT reactions
involving heavy projectile and target combinations have been extensively studied
experimentally near barrier energies over the last few
years~\cite{kozulin2012,kozulin2014,kratz2013,watanabe2015,devaraja2015,desai2019,birkenbach2015,vogt2015,galtarossa2018,kalandarov2020,li2020b,adamian2021,heinz2022,wu2022}.
In quasi-fission reactions involving heavy nuclei, ions stick together to form a
dinuclear system. During this period, a large number of nucleons are transferred
between the colliding ions. It has been experimentally observed that entrance
channel properties such as N/Z ratio, the presence of entrance-channel magicity
(the number of spherical shells in the reaction entrance
channel)~\cite{simenel2012b}, relative orientation of deformed
ions~\cite{hinde1996,knyazheva2007,nishio2008,ghosh2004,wakhle2014}, and the charge
product Z$_P$Z$_T$~\cite{back1996,hinde2002,rafiei2008,kozulin2019,itkis2011}
could have a significant effect on the reaction outcome. Although quasi-fission
reactions usually favor transfer towards mass symmetry, it has been experimentally observed
that nucleon transfer from lighter to heavier ions is also
possible~\cite{kozulin2014,kozulin2017,watanabe2015}.

Exploring the mechanisms
behind multinucleon transfer reactions requires a comprehensive theoretical
framework that can properly account for these intricate processes. The time-dependent
Hartree-Fock (TDHF) theory provides a microscopic description of reaction
dynamics and it has been employed extensively to analyze MNT
reactions~\cite{umar2016,hammerton2015,nakatsukasa2016,oberacker2014,umar2015a,sekizawa2016,sekizawa2017,simenel2010,simenel2020} (see~\cite{simenel2012,simenel2018,sekizawa2019,stevenson2019}
for recent reviews of TDHF applications to heavy-ion reactions). In TDHF
theory, it is possible to calculate the physical dynamical observables, such as
the mean values of the fragment charge and mass, and the mean kinetic energy depletion
arising from one body dissipation~\cite{ayik2020b}. Although the TDHF theory has shown that it is a
strong candidate for exploring MNT reactions, it is unable to account for
fluctuations and dispersions of the fragment mass and charge distributions. To
overcome this limitation, one has to go beyond the mean-field
approximation~\cite{tohyama2002a,tohyama2020,simenel2011,ayik2008,lacroix2014}.
One such approach is through the time-dependent random phase approximation
(TDRPA) developed by Balian and V\'en\'eroni, which offers a reasonable theory
to compute larger observable fluctuations beyond mean-field. This method
has been used to study multinucleon transfer reactions in symmetric
systems~\cite{balian1985,balian1992,williams2018,godbey2020,godbey2020b}, and
was validated by comparing to experimental data.
However, the method
is limited to describe the dispersion of charge and mass distributions only
for symmetric systems.

In analogy with TDRPA, the stochastic mean-field (SMF) theory, which was first
proposed by Ayik in 2008~\cite{ayik2008}, provides an alternative extension to the
mean-field approximation. The SMF theory goes beyond the TDHF method by including
mean-field fluctuations and correlations into the description. This work aims to
show that the quantal diffusion approach based on SMF theory can explain the
reaction dynamics observed in $^{48}$Ca+$^{244}$Pu reaction at
E$_\text{{c.m.}}=$~203.2~MeV and $^{86}$Kr+$^{198}$Pt reaction at E$_\text{{c.m.}} = $~324.2~MeV
without any adjustable parameters other than standard Skyrme
energy density functional parameters. The remainder of this paper is organized
as follows. In Sec.~\ref{sec_2}, we present TDHF calculations for
$^{48}$Ca+$^{244}$Pu reaction at E$_\text{{c.m.}}= $~203.2~MeV and
$^{86}$Kr+$^{198}$Pt reaction at E$_\text{{c.m.}} = $~324.2~MeV. In
Sec.~\ref{sec_3}, we briefly discuss the quantal transport theory based on the
SMF approach and discuss the SMF results for both systems. In Sec.~\ref{sec_4}, we
present the primary product mass distributions for both reactions
 using the quantal diffusion approach based on the SMF
theory and compare our results with the available experimental data~\cite{kozulin2019,sen2022}.
In Sec.~\ref{sec_5}, conclusions are presented.

\section{Mean reaction dynamics in TDHF}\label{sec_2}

\begin{table}[!t]
\centering
\caption{Results of the TDHF calculations for the
${}^{48}\text{Ca}+{}^{244}\text{Pu}$ system at $E_{\text{c.m.}}=203.2$~MeV for the
tip orientation of the ${}^{244}\text{Pu}$ nucleus and the
${}^{86}\text{Kr}+{}^{198}\text{Pt}$ system at $E_{\text{c.m.}}=324.2$~MeV for the
tip orientation of the ${}^{198}\text{Pt}$ nucleus.}
\begin{ruledtabular}
${}^{48}\text{Ca}+{}^{244}\text{Pu}$ system at $E_{\text{c.m.}}=203.2$~MeV
\begin{tabular}{c c c c c c c c c c }
$\ell_i$ $(\hbar)$ &$A_1^f$ &$Z_1^f$  &$A_2^f$ & $Z_2^f$  & $TKE$   & $\theta_{cm}$ & $\theta_{1}^{lab}$ & $\theta_{2}^{lab}$  \\
\hline
 44 &88.9 &35.4 &203.1 &78.6 &218.8 &94.9 &78.8 &53.7\\
 46 &88.4 &35.1 &203.6 &78.9 &217.8 &89.1 &73.4 &57.6\\
 48 &89.4 &35.0 &203.6 &79.0 &216.6 &84.0 &68.6 &61.4\\
 50 &88.2 &35.0 &203.8 &79.0 &217.2 &79.7 &64.9 &64.3\\
 52 &87.7 &34.9 &204.3 &79.1 &218.3 &76.7 &62.4 &66.5\\
 54 &87.6 &34.9 &204.4 &79.1 &218.4 &74.1 &60.1 &68.5\\
 56 &87.9 &35.1 &204.1 &78.9 &216.9 &70.9 &57.2 &71.0\\
 58 &86.8 &34.7 &205.2 &79.3 &215.2 &66.9 &53.8 &73.6\\
 60 &83.8 &35.5 &208.2 &80.5 &214.8 &62.2 &50.1 &76.4\\
 62 &88.7 &35.5 &203.3 &78.5 &215.6 &56.3 &44.7 &83.4\\
 64 &92.1 &36.6 &199.9 &77.4 &211.1 &43.6 &34.1 &-83.1\\
 66 &84.5 &33.9 &207.5 &80.1 &197.5 &43.7 &34.4 &-88.5\\
 68 &80.8 &32.5 &211.2 &81.5 &190.0 &57.4 &45.7 &76.4\\
 70 &65.0 &26.5 &227.0 &87.5 &168.2 &75.1 &61.9 &55.9\\
 \hline
\multicolumn{9}{c}{${}^{86}\text{Kr}+{}^{198}\text{Pt}$ system at $E_{\text{c.m.}}=324.2$~MeV} \\
\hline
 60 &95.9 &39.1 &188.1 &74.9 &214.2 &109.4 &75.3 &32.8\\
 80 &89.7 &36.7 &194.3 &77.3 &220.6 &102.9 &71.8 &34.9\\
 100 &88.4 &36.2 &195.6 &77.8 &223.6 &98.4 &68.6 &36.7\\
 120 &88.3 &36.2 &195.7 &77.8 &242.5 &93.9 &66.0 &39.7\\
\end{tabular}
\end{ruledtabular}
\label{tab_ca+pu_kr+pt_tdhf}
\end{table}
\begin{figure}[!htb]
  \centering
  \includegraphics[width=8.6cm]{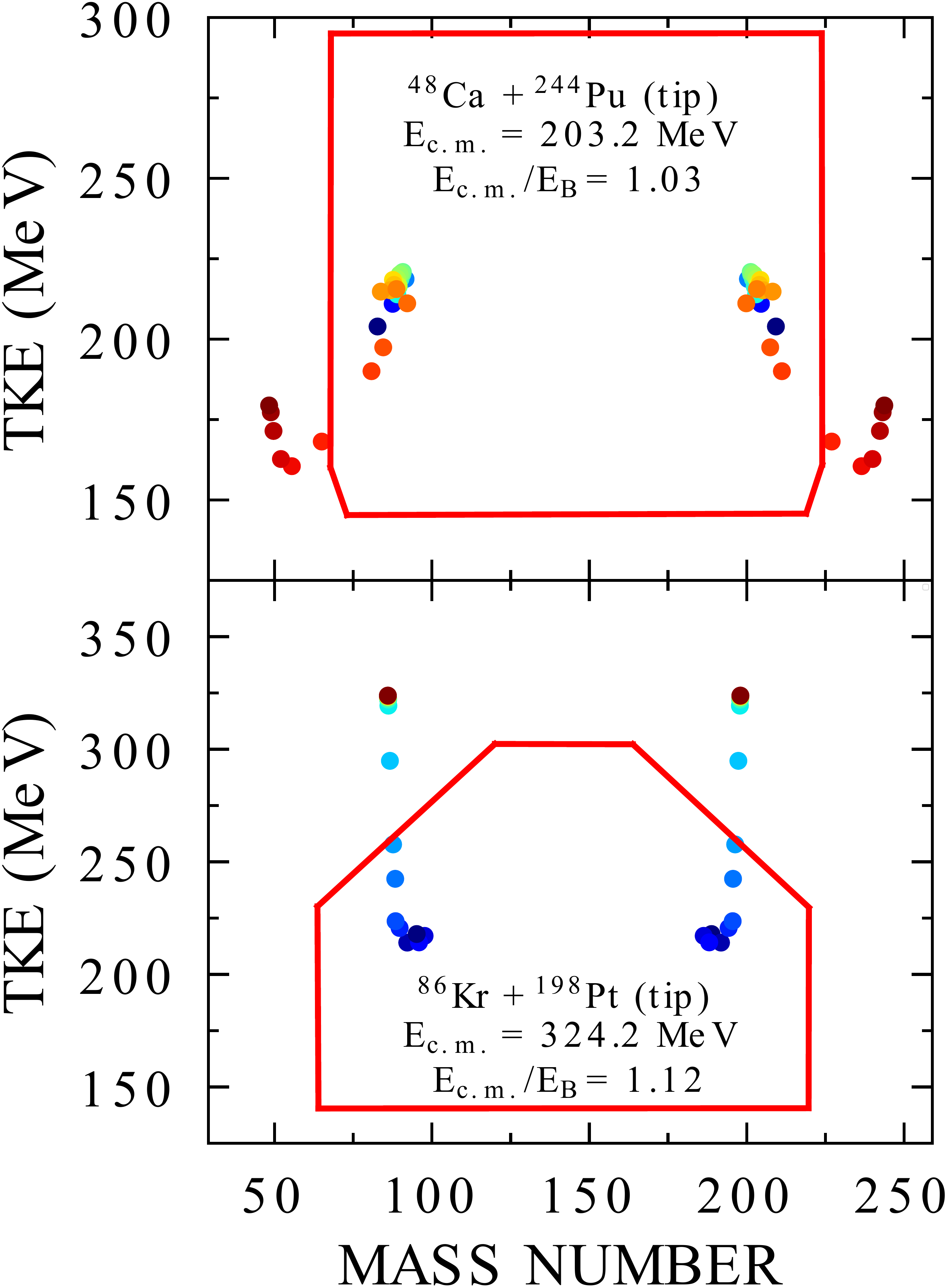}
  \caption{Mass-energy distributions of fragments obtained by TDHF calculations
for the ${}^{48}\text{Ca}+{}^{244}\text{Pu}$ system at
$E_{\text{c.m.}}=203.2$~MeV and the ${}^{86}\text{Kr}+{}^{198}\text{Pt}$ system
at $E_{\text{c.m.}}=324.2$~MeV. Red continuous contours represent the gates used
in the experiments in Refs.~\cite{kozulin2014c,kozulin2019,sen2022} }
  \label{fig_masstke_combined}
\end{figure}
In Table ~\ref{tab_ca+pu_kr+pt_tdhf}, we share the TDHF results for
${}^{48}\text{Ca}+{}^{244}\text{Pu}$ system at $E_\text{{c.m.}}=203.2$~MeV for the
tip orientation of the ${}^{244}\text{Pu}$ nucleus and
${}^{86}\text{Kr}+{}^{198}\text{Pt}$ system at $E_{\text{c.m.}}=324.2$~MeV for the
tip orientation of the ${}^{198}\text{Pt}$ nucleus for a range initial
orbital angular momenta, $\ell_i$, final values of mass and charge numbers of
projectile-like $A_1^f$, $Z_1^f$ and target-like $A_2^f$, $Z_2^f$ fragments,
final total kinetic energy TKE, scattering angles in the center of
mass frame $\theta_{\text{c.m.}}$, and laboratory frame $\theta_1^{lab}$ and
$\theta_2^{lab}$, respectively. The calculations presented in the article
employed the TDHF code~\cite{umar1991a,umar2006c} using the SLy4d Skyrme energy
density functional~\cite{kim1997}, with a box size of $60\times 60\times 36$~fm
in the $x-y-z$ directions. To reduce the computation time, we present quantities
that are evaluated for every two units of initial angular momentum for
${}^{48}\text{Ca}+{}^{244}\text{Pu}$ system, and 20 units of initial angular
momentum for ${}^{86}\text{Kr}+{}^{198}\text{Pt}$ system. TDHF ground state
calculations show that $^{48}$Ca and $^{86}$Kr nuclei exhibit spherical shapes,
whereas $^{244}$Pu and $^{198}$Pt nuclei exhibit strong prolate and oblate
deformations, respectively. For this reason, to consider all possible relative
orientations of the deformed nuclei, TDHF and SMF calculations were performed
for all possible relative orientations of the reaction partners. As a
convention, we denote the initial orientation of the target principal
deformation axis to be in the beam direction as the tip, and the case when
their principal axis is perpendicular to the beam direction as the side.
\begin{figure}[!htb]
  \centering
  \includegraphics[width=8.6cm]{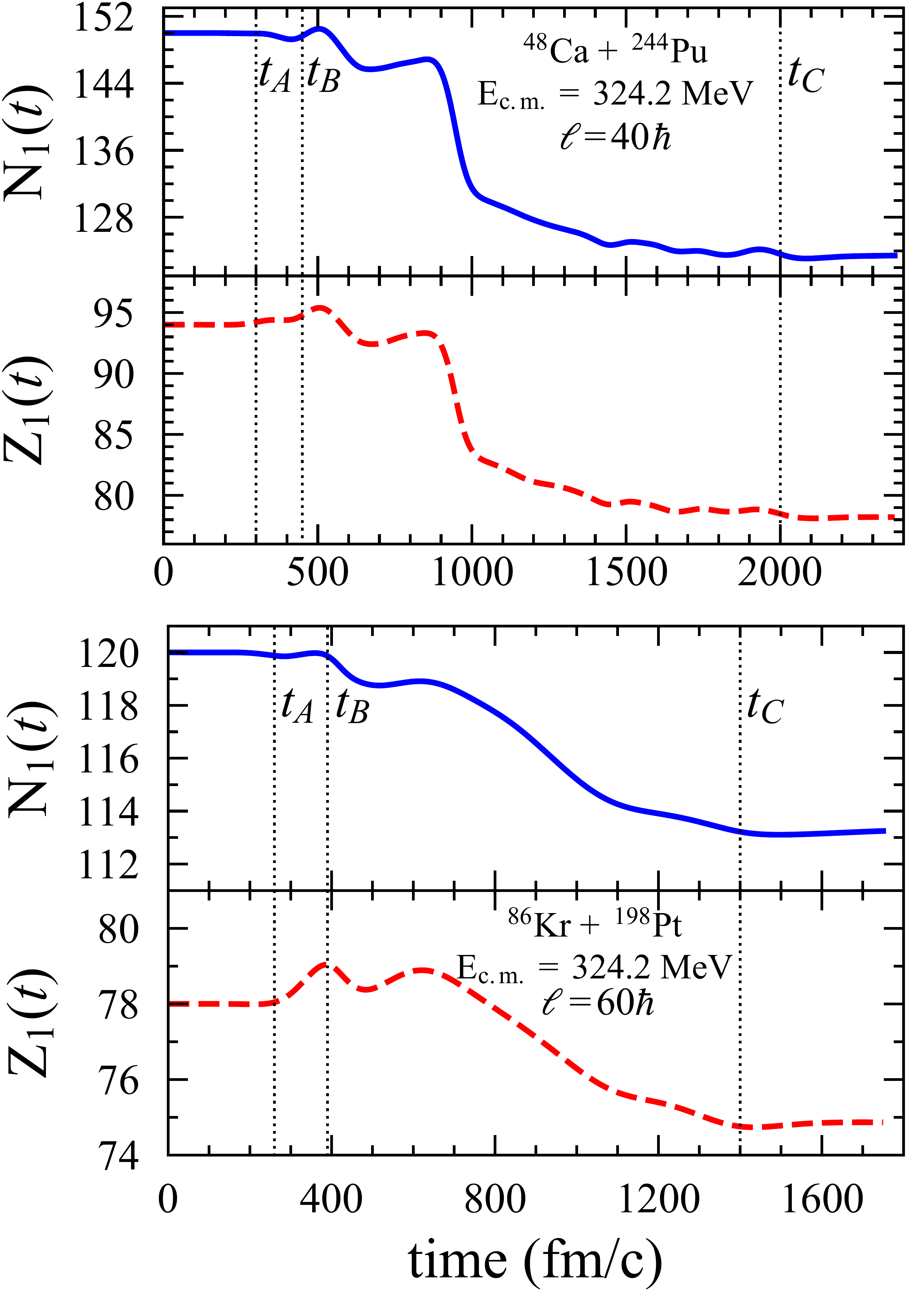}
  \caption{Mean values of neutron and proton numbers of target-like fragments at
initial angular momentum $\ell_i=40\hbar$ for
${}^{48}\text{Ca}+{}^{244}\text{Pu}$ system at $E_{\text{c.m.}}=203.2$~MeV and
$\ell_i=60\hbar$ for ${}^{86}\text{Kr}+{}^{198}\text{Pt}$ system at
$E_{\text{c.m.}}=324.2$~MeV are shown as a function of time. Solid blue lines
denote the neutron numbers and dashed red lines denote the proton numbers of
target-like fragments. The labels $t_A$, $t_B$, and $t_C$ indicate the
projection of the time intervals used to determine the curvature parameters. }
  \label{fig_nzt}
\end{figure}

In Ref.~\cite{kozulin2014c}, ${}^{48}\text{Ca}+{}^{244}\text{Pu}$ system is
analyzed at two different energies, $E_{\rm{c.m.}}=193.3$~MeV and
$E_{\rm{c.m.}}=203.2$~MeV, corresponding to $E_\text{c.m.} /V_\text{Bass} = 0.98
$ and $E_\text{c.m.} /V_\text{Bass} = 1.03 $, Similarly in
Refs.~\cite{sen2022,kozulin2019}, ${}^{86}\text{Kr}+{}^{198}\text{Pt}$ system is
analyzed at $E_{\rm{c.m.}}=324.2$~MeV corresponding to $E_\text{c.m.}
/V_\text{Bass} = 1.12 $.  In these experiments, the reaction fragments for
${}^{48}\text{Ca}+{}^{244}\text{Pu}$ and ${}^{86}\text{Kr}+{}^{198}\text{Pt}$
systems were detected between $42^\circ-78^\circ$ and $30^\circ-68^\circ$ in the
laboratory frame. The mass-energy distributions of the detected fragments are
shown in Fig.~2 of Ref.~\cite{kozulin2019} and Fig.~1 of Ref.~\cite{sen2022}.
In these figures, red continuous contours were used to exclude elastic events
and select events arising from either MNT or fission-like reactions. In order to
determine the initial angular momenta intervals corresponding to the
experimental windows, the TKE vs. mass value is
scattered in the TKE-mass plane for each calculated initial angular momentum value and the
resultant distribution is plotted for both systems in
Fig.~\ref{fig_masstke_combined}. Red continuous contours represent the gates
used in the experiments in Refs.~\cite{kozulin2014c,kozulin2019,sen2022}. By
considering the mean field calculations shown in Fig.~\ref{fig_masstke_combined} and the scattering angles in the laboratory frame,
$\theta^{lab}_1$ and $\theta^{lab}_2$, given in Table~\ref{tab_ca+pu_kr+pt_tdhf}, we determined that the initial angular momenta range to be $44\hbar \leq \ell_i \leq 70\hbar~$ for the
${}^{48}\text{Ca}+{}^{244}\text{Pu}$ system and $60\hbar \leq \ell_i \leq
120\hbar~$ for the ${}^{86}\text{Kr}+{}^{198}\text{Pt}$ system, corresponding to the
angular coverage range in these experiments~\cite{kozulin2014c,kozulin2019,sen2022}.
For the side orientation of the target nuclei, we observed that the primary binary
products do not satisfy the angular range and TKE-mass region conditions
related to the experimental setup mentioned above at the same time. Thus, in
this study, calculations for side orientations were excluded.
\begin{figure}[!htb]
  \centering
  \includegraphics[width=8.6cm]{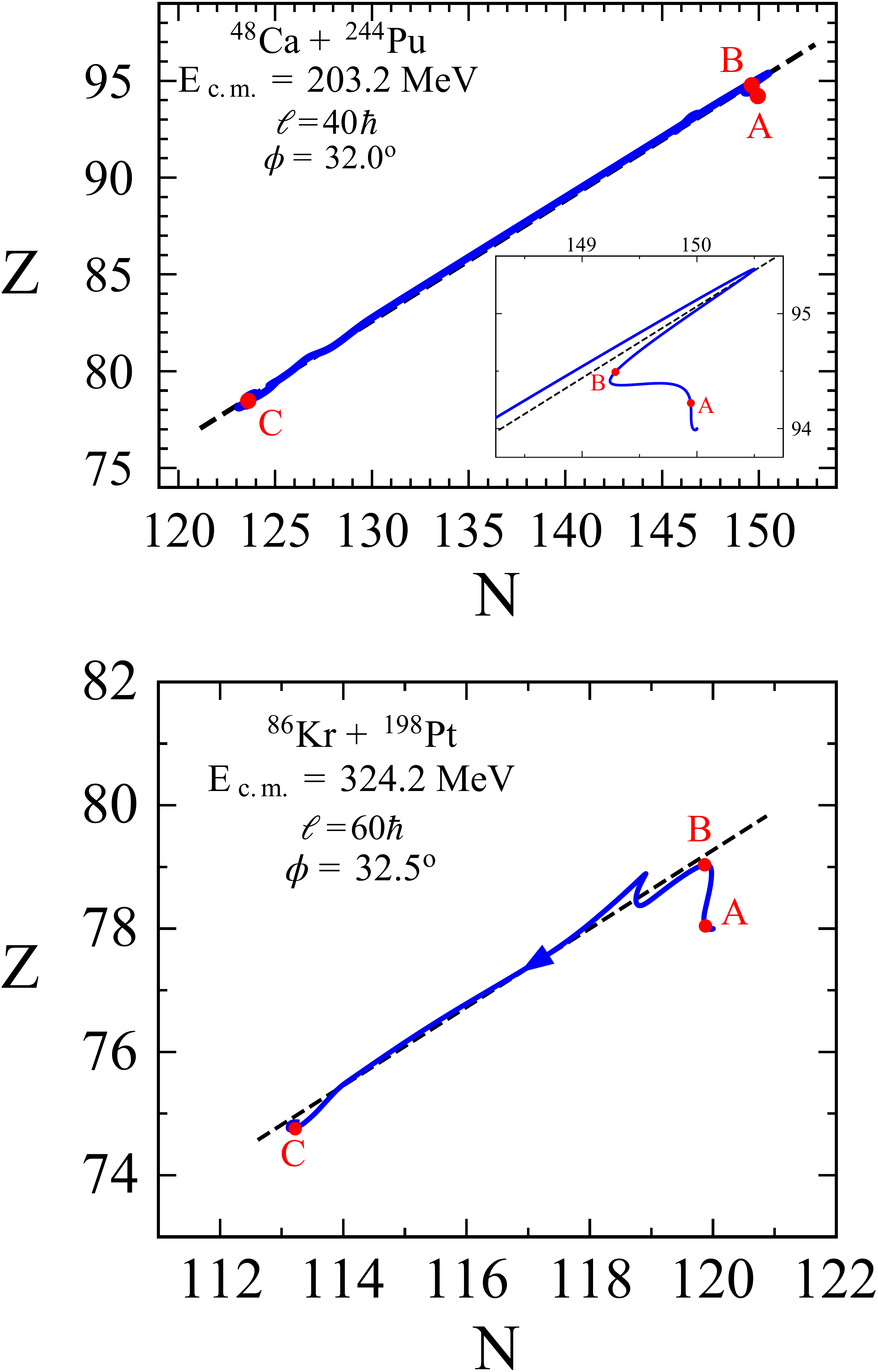}
  \caption{Mean drift path in the $N-Z$ plane for the target-like fragments are
given at initial angular momentum $\ell_i=40\hbar$ for
${}^{48}\text{Ca}+{}^{244}\text{Pu}$ system at $E_{\text{c.m.}}=203.2$~MeV and
$\ell_i=60\hbar$ for ${}^{86}\text{Kr}+{}^{198}\text{Pt}$ system at
$E_{\text{c.m.}}=324.2$~MeV. Solid blue lines denote the mean drift path and
dashed black lines denote the iso-scalar lines. The labels A, B, and C indicate
the projection of the time intervals used to determine the curvature
parameters.  }
  \label{fig_nz}
\end{figure}

If we denote the charge asymmetry of the reaction fragments with
$\delta=\frac{N-Z}{A}$, the charge asymmetry values of the reaction partners are
found to be $\delta(\mathrm{Ca}) \simeq 0.17 $, $\delta(\mathrm{Pu}) \simeq 0.23 $,
$\delta(\mathrm{Kr}) \simeq 0.16$, and $\delta(\mathrm{Pt}) \simeq 0.21 $. When reaction
partners have different charge asymmetries, particularly during the initial phase of
the collision, reaction partners quickly exchange few nucleons to reach
the same charge asymmetry value. Subsequently, the reaction partners continue to exchange
nucleons and the system drifts along the nearly constant charge asymmetry line
in the $(N-Z)$ plane, which we call the iso-scalar line. As an example of the
reaction dynamics calculated within TDHF framework, we present the time
evolution of the neutron $N(t)$ and proton $Z(t)$ numbers of target-like
fragments in ${}^{48}\text{Ca}+{}^{244}\text{Pu}$ at $\ell_i = 40\hbar$ and
${}^{86}\text{Kr}+{}^{198}\text{Pt}$ system at $\ell_i = 60\hbar$ in
Fig.~\ref{fig_nzt}. To gain more insight into these systems, we can easily
eliminate the time dependence of neutron $N(t)$ and proton $Z(t)$ numbers and
plot the collision dynamics in the $(N-Z)$ plane. In the resultant
Fig.~\ref{fig_nz}, thick blue lines show the mean drift path of the target-like
fragments, whereas the dashed lines show the iso-scalar lines corresponding to
each system. The angle $\phi$ between the N-plane and iso-scalar path was found
to be $\phi=32.0^\circ$ and $\phi=32.5^\circ$ for each system. In both systems,
the iso-scalar line extends from the lighter reaction partner up to a heavier
reaction partner, passing through the mass symmetry point $(N_0, Z_0)$, where
$N_0=(N_1+N_2)/2 $. As we can clearly see in Fig.~\ref{fig_nz}, because the
charge asymmetries of the initial reaction partners differ in both systems, both
systems initially quickly drift towards the iso-scalar line. The reaction
partners then drift along the stability line toward the mass symmetry point. In
Fig.~\ref{fig_nzt} and Fig.~\ref{fig_nz}, labels $t_A$, $t_B$, and $t_C$
indicate the time labels used in calculating the reduced curvature parameters
which will be explained in Sec.~\ref{sec_3c}. Lastly, within the given initial
angular momentum intervals, the mean reaction times are roughly equal to 6~zs for the
${}^{48}\text{Ca}+{}^{244}\text{Pu}$ system, and 3~zs for the
${}^{86}\text{Kr}+{}^{198}\text{Pt}$ system, which is consistent with the
predictions in Refs.~\cite{shen1987,kozulin2019}.

\section{Quantal Diffusion Description}\label{sec_3}

\subsection{Langevin equation for macro variables}\label{sec_3a}

In the TDHF approach, with a given set of initial conditions, a single-particle
density matrix is calculated by a single Slater determinant.
\begin{comment}
In the SMF approach, the correlated initial state is represented by an ensemble
of single-particle density matrices which are specified in terms of initial
fluctuations.
\end{comment}
In the SMF approach, the collision dynamics are described in terms of an
ensemble of the mean-field events~\cite{supplemental,ayik2008,lacroix2014}. In
each event, the complete set of single-particle wave functions is determined by
the TDHF equations with the self-consistent Hamiltonian of that event. The
ensemble is considered to be generated by incorporating the quantal and thermal
fluctuations at the initial state. We consider low-energy collisions at
which a di-nuclear structure is maintained during the collision. The identities of
colliding nuclei are preserved to a large extent, but nucleon transfer takes
place between target-like and projectile-like nuclei. With the help of geometric
projection, for small amplitude fluctuations, we derive the linearized coupled
Langevin equations (Eq.~(7) in Ref.~\cite{supplemental}) for macroscopic variables. For
further explanation of Langevin equation for macroscopic variables in nuclear
reactions, we refer the reader to Sec.~1A in the supplementary online
material~\cite{supplemental}.

\subsection{Quantal Diffusion Coefficients}\label{sec_3b}

According to the SMF approach, stochastic parts of drift coefficients (given by
Eq.~(8) in Ref.~\cite{supplemental}) have Gaussian random distributions with zero
mean values, $\delta \bar{v}_{p}^{\lambda}(t)=0,~\delta
\bar{v}_{n}^{\lambda}(t)=0$ with the autocorrelation functions, that are
integrated over the history, determining the diffusion coefficients $D_{\alpha
\alpha}(t)$ for proton and neutron transfers. Usually, to calculate the quantal
diffusion coefficients,  one needs to sum over the whole particle and hole
states. But in the diabatic limit, we can get rid of the particle states by
utilizing closure relations and calculate the diffusion coefficients only in
terms of the occupied single-particle states of the TDHF calculations. This is
consistent with the dissipation-fluctuation theorem of non-equilibrium statistical
mechanics. The explicit expression for the quantal diffusion coefficients is
given in Eq.~(10) in the supplementary online material~\cite{supplemental}. Such a
quantal expression for the nucleon diffusion coefficient in heavy-ion collisions
is given in the literature for the first time from a microscopic basis.
\begin{figure}[!htb]
  \centering
  \includegraphics[width=8.6cm]{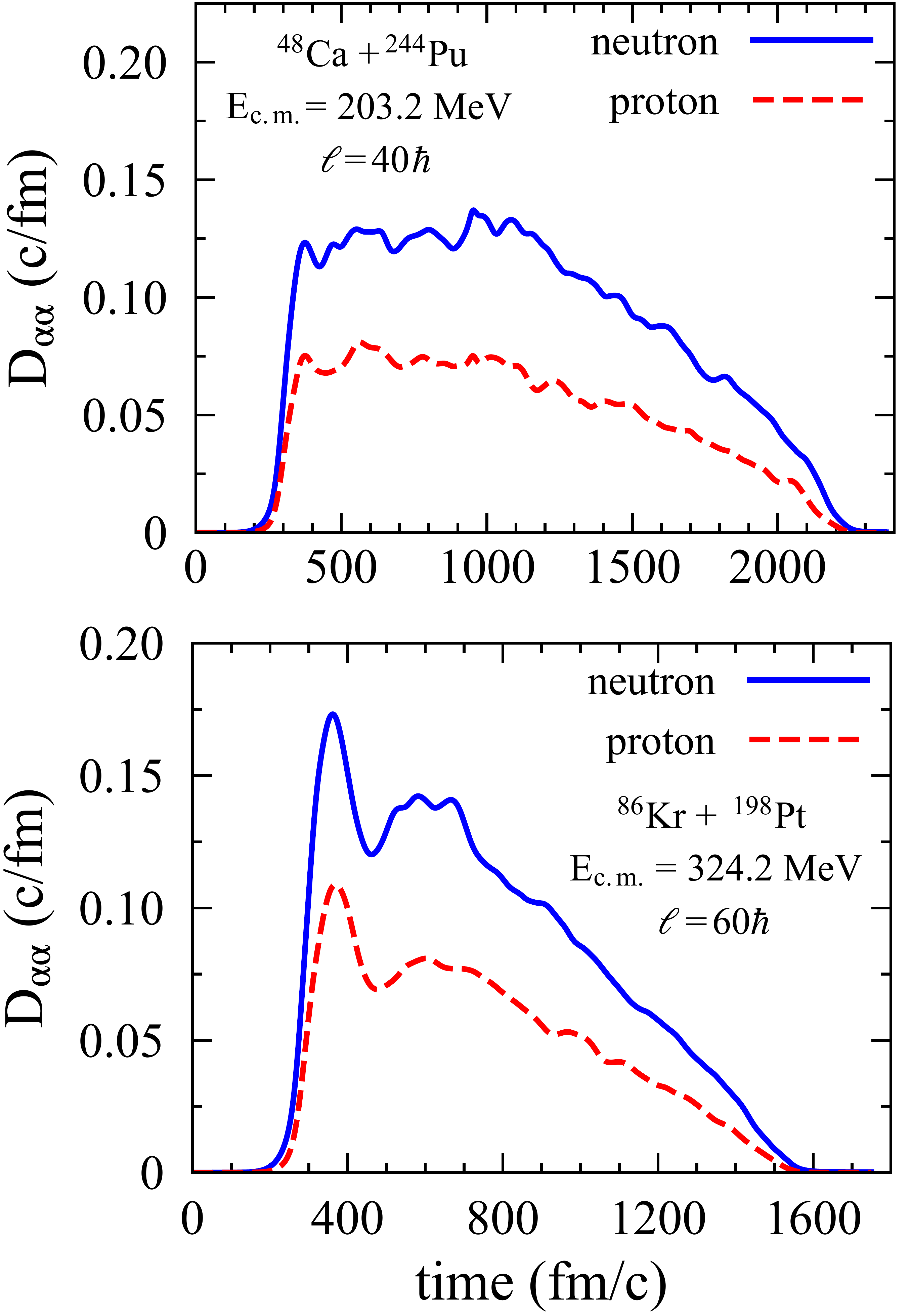}
  \caption{Diffusion coefficient for neutron and proton transfers at initial
angular momentum $\ell = 40\hbar$ for ${}^{48}\text{Ca}+{}^{244}\text{Pu}$
system at $E_{\text{c.m.}}=203.2$~MeV, and $\ell = 60\hbar$ for
${}^{86}\text{Kr}+{}^{198}\text{Pt}$ system at $E_{\text{c.m.}}=324.2$~MeV are
shown as a function of time. Solid blue lines denote the diffusion coefficients
of neutron transfer, $D_{NN}(t)$; dashed red lines denote the diffusion
coefficients of proton transfer, $D_{ZZ}(t)$. }
  \label{fig_diff}
\end{figure}

In Fig.~\ref{fig_diff}, the diffusion coefficients for neutron and proton
transfers are shown. The quantal diffusion coefficients were calculated using
Eq.~(10) in supplementary online material~\cite{supplemental} (see also Refs.~\cite{gardiner1991,merchant1982}),
for each initial angular momentum.  As shown
in the figure, the neutron diffusion coefficients are almost twice the value of
the proton diffusion coefficients. This is mainly due to the Coulomb repulsion
between the protons and the total neutron number of the colliding system being
far greater than the total proton number. This behavior is also observed in
different heavy-ion reactions in which the SMF theory is
employed~\cite{sekizawa2020,ayik2023,ayik2021,ayik2020,ayik2019,ayik2019b,yilmaz2018,yilmaz2020}.

\subsection{Reduced Curvature Parameters}\label{sec_3c}

After the colliding nuclei form a di-nuclear system, the nucleon drift mechanism is
mainly determined via the potential energy surface projected in the $(N-Z)$
plane.  TDHF theory includes different energy contributions microscopically such
as surface energy, electrostatic energy, symmetry energy, and centrifugal
potential energy. In SMF theory, we approximate the shape of the potential energy in the
$(N-Z)$ plane near the local equilibrium state in terms of two parabolic forms,
given by Eq.~(18) in Ref.~\cite{supplemental}. By using Einstein relations in the
overdamped limit~\cite{ayik2017,ayik2018,yilmaz2018,ayik2019,sekizawa2020} and
inverting Eq.~(17) of Ref.~\cite{supplemental}, we can derive Eqs.~(25-28) to
calculate reduced curvature parameters.

As we briefly mentioned in Sec.~\ref{sec_2}, to calculate the reduced curvature
parameters related to each system, we considered the collision at $\ell_i =
40\hbar$ for ${}^{48}\text{Ca}+{}^{244}\text{Pu}$ system and collision at
$\ell_i = 60\hbar$ for ${}^{86}\text{Kr}+{}^{198}\text{Pt}$ system. By using
Eqs.~(25-28) in Ref.~\cite{supplemental}, we can calculate the iso-scalar and
iso-vector reduced curvature parameters related to each system. In Table~\ref{tab3},
we tabulate the calculated values of reduced curvature parameters and
time intervals related to each system. The local equilibrium state $(N_0,Z_0)$
in Eqs.~(26,28) in Ref.~\cite{supplemental} is taken as mass symmetry point
$(N_0,Z_0) = (N_{T}/2,Z_{T}/2)$ for both systems; where its value is equal to
$(N_0,Z_0)=(89,57)$ for the ${}^{48}\text{Ca}+{}^{244}\text{Pu}$ system and
$(N_0,Z_0)=(85,57)$ for the ${}^{86}\text{Kr}+{}^{198}\text{Pt}$ system.

\begin{table}[!htb]
\caption{Calculated reduced curvature parameters for
${}^{48}\text{Ca}+{}^{244}\text{Pu}$ system at $E_{cm}=203.2$~MeV and
${}^{86}\text{Kr}+{}^{198}\text{Pt}$ system at $E_{\text{c.m.}}=324.2$~MeV. The
time intervals $t_A$ and $t_B$ are used to calculate the iso-vector reduced
curvature parameter $\alpha$ using Eqs.~(27,28) of Ref.~\cite{supplemental}.
The time intervals $t_B$ and $t_C$ are used to calculate the isoscalar-reduced
curvature parameter, $\beta$, using Eqs.~(25,26) of Ref.~\cite{supplemental}.
These time labels are also shown in Fig.~(\ref{fig_nz}) and
Fig.~(\ref{fig_nzt}).}
\label{tab3}
\begin{ruledtabular}
\begin{tabular}{ c | c | c | c | c | c  }
System &$t_A(fm/c)$ &$t_B(fm/c)$ &$t_C(fm/c)$ &$\alpha$ &$\beta$\\
\hline
${}^{48}\text{Ca}+{}^{244}\text{Pu}$ &300 &425 &2000 &0.156 &0.004 \\
\hline
${}^{86}\text{Kr}+{}^{198}\text{Pt}$ &260 &390 &1400 &0.176 &0.003 \\
\end{tabular}
\end{ruledtabular}
\end{table}

\subsection{Covariances of fragment charge and mass distributions}\label{sec_3d}
\begin{figure}[!htb]
  \centering
  \includegraphics[width=8.6cm]{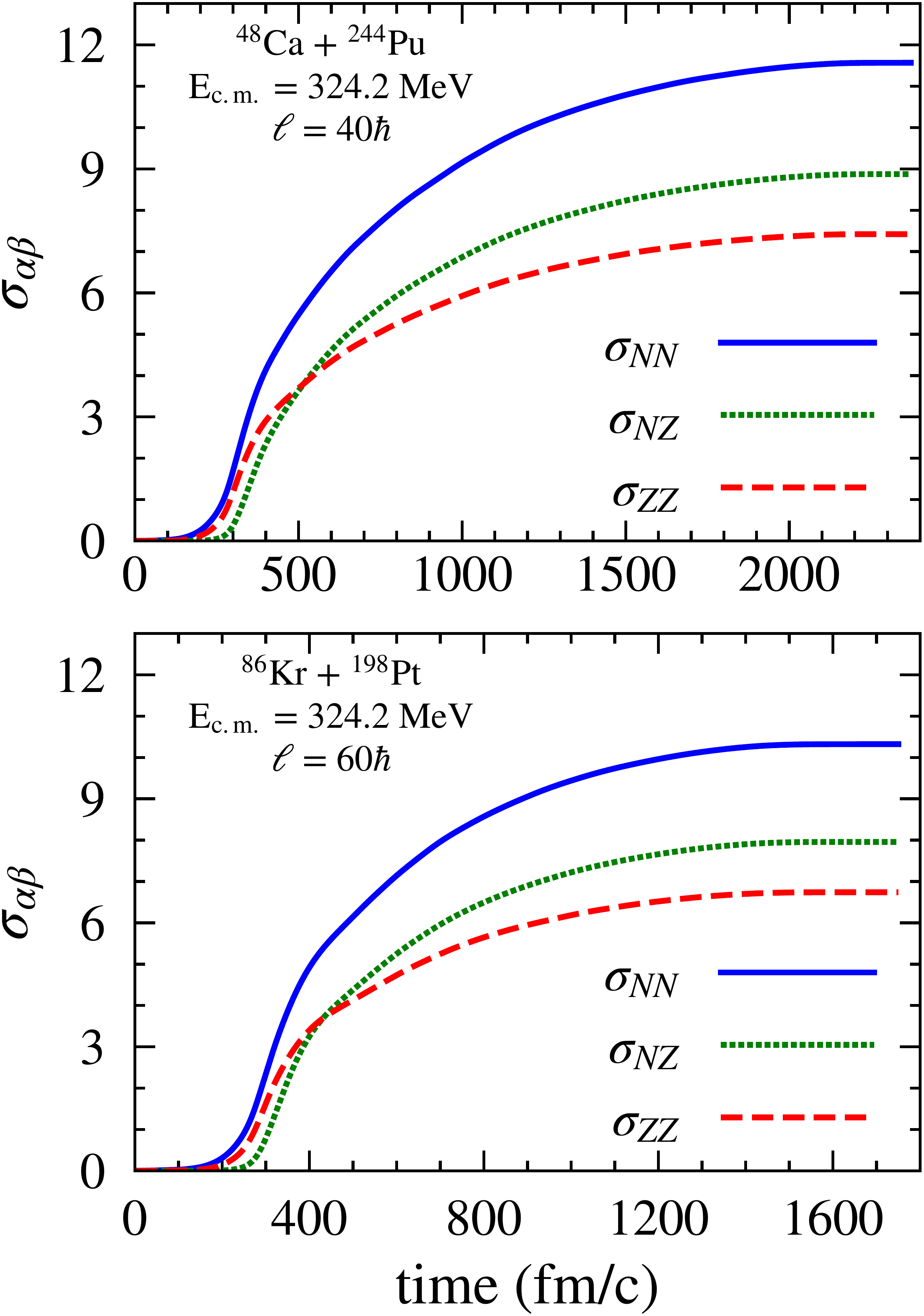}
  \caption{Neutron, proton, and mixed variances as a function of time at initial
angular momentum $\ell = 40\hbar$ for ${}^{48}\text{Ca}+{}^{244}\text{Pu}$
system at $E_{\text{c.m.}}=203.2$~MeV, and $\ell = 60\hbar$ for
${}^{86}\text{Kr}+{}^{198}\text{Pt}$ system at $E_{\text{c.m.}}=324.2$~MeV.
Solid blue lines denote the neutron variances, $\sigma_{NN}$; dashed red lines
denote the proton variances, $\sigma_{ZZ}$; and dotted green lines denote the
mixed variances, $\sigma_{NZ}$.}
  \label{fig_disp}
\end{figure}

Primary mass distribution is found by using the standard expression
\begin{align}\label{eq_1}
Y(A)^{MNT}=\frac{1}{\sum_{\ell_{\min }}^{\ell_{\max }}(2 l+1)} \sum_{\ell_{\min }}^{\ell_{\max }}(2 \ell+1) P_{\ell}(A)\;.
\end{align}
Here, $ P_{\ell}(A)$ is given by
\begin{equation}\label{eq_2}
P_{\ell}(A)=\frac{1}{2}\left[P_{\ell}^{pro}(A)+P_{\ell}^{tar}(A)\right]\;,
\end{equation}
where $P_{\ell}^{pro}(A)$ and $P_{\ell}^{tar}(A)$ denote the normalized
probabilities of producing projectile-like and target-like fragments,
respectively. A factor of $1/2$ was introduced to normalize the total primary
fragment distribution to unity. In Eq.~(\ref{eq_2}), while probabilities are given by
\begin{align}\label{eq_3}
P_{\ell}^{pro}(A)=\frac{1}{\sqrt{2 \pi}} \frac{1}{\sigma_{AA}^{2}(\ell)} \exp \left[-\frac{1}{2}\left(\frac{A-A_{\ell}^{pro}}{\sigma_{AA}^{2}(\ell)}\right)^{2}\right]\;.
\end{align}
Here, the probability distribution of the mass number of produced fragments is
determined by summing over N or Z and keeping the total mass number constant A =
N + Z in correlated Gaussian function Eq.~(29) in Ref.~\cite{supplemental}. In
Eq.~(\ref{eq_3}), the mass variance is given by $\sigma_{A A}^{2}(\ell)=\sigma_{N
N}^{2}(\ell)+\sigma_{Z Z}^{2}(\ell)+2 \sigma_{N Z}^{2}(\ell)$, where $\sigma_{N
N}^{2}(\ell)$, $\sigma_{Z Z}^{2}(\ell)$, and $\sigma_{N Z}^{2}(\ell)$ stand for
neutron, proton, and mixed dispersions, respectively. By using the calculated
reduced curvature parameters and quantal diffusion coefficients for proton and
neutron transfer, we can solve the coupled differential equations (Eqs.~(14-16)
of Ref.~\cite{supplemental} and Refs.~\cite{schroder1981,weiss1999}) to determine $\sigma^2_{NN}, \sigma^2_{ZZ},
\sigma^2_{NZ}$, and $\sigma^2_{AA}$ for each initial angular momentum with the
initial conditions $\sigma_{NN} =\sigma_{ZZ} =\sigma_{NZ} = 0 $ at $t=0$. In
Tab.~\ref{tab_ca+pu_kr+pt_smf}, we show the asymptotic values for
neutrons $\sigma_{NN}$, protons $\sigma_{ZZ}$, mixed dispersions $\sigma_{NZ}$
and mass dispersions $\sigma_{AA}$ calculated for each initial angular momentum
value. As an example, the calculated neutron, proton, and mixed variances as a
function of time in ${}^{48}\text{Ca}+{}^{244}\text{Pu}$ reaction at $\ell_i =
40\hbar$ and ${}^{86}\text{Kr}+{}^{198}\text{Pt}$ reaction at $\ell_i = 60\hbar$
are shown in Fig~\ref{fig_disp}. In both systems, we see that during the initial phase
of the reaction, up to about $t\simeq500(fm/c)$, the magnitude of variances are in order as
$\sigma_{NZ}<\sigma_{ZZ}<\sigma_{NN}$. After that the correlations evolve over time, changing
the order to $\sigma_{ZZ} < \sigma_{NZ}< \sigma_{NN}$, demonstrating the
importance of correlations arising from significant energy dissipation.
\begin{table}[!htb]
    \centering
    \caption{Results of the SMF calculations for the
        ${}^{48}\text{Ca}+{}^{244}\text{Pu}$ system at $E_{\text{c.m.}}=203.2$~MeV in
        tip configuration of ${}^{244}\text{Pu}$ nucleus and the
        ${}^{86}\text{Kr}+{}^{198}\text{Pt}$ system at $E_{\text{c.m.}}=324.2$~MeV in
        tip configuration of ${}^{198}\text{Pt}$ nucleus.}
    \begin{ruledtabular}
        ${}^{48}\text{Ca}+{}^{244}\text{Pu}$ system at $E_{\text{c.m.}}=203.2$~MeV
        \begin{tabular}{c  c c c c }
            $\ell_i$ $(\hbar)$  & $\sigma_{NN}$ & $\sigma_{ZZ}$ & $\sigma_{NZ}$ & $\sigma_{AA}$ \\
            \hline
            44 &11.6 &7.4 &8.9 &18.7 \\
            46 &11.6 &7.5 &8.9 &18.7 \\
            48 &11.7 &7.5 &9.0 &18.8 \\
            50 &11.7 &7.5 &9.0 &18.8 \\
            52 &11.7 &7.5 &8.9 &18.8 \\
            54 &11.6 &7.4 &8.9 &18.6 \\
            56 &11.5 &7.4 &8.8 &18.5 \\
            58 &11.4 &7.3 &8.8 &18.4 \\
            60 &11.5 &7.4 &8.8 &18.5 \\
            62 &11.5 &7.4 &8.8 &18.5 \\
            64 &11.9 &7.6 &9.2 &19.2 \\
            66 &11.5 &7.4 &8.8 &18.5 \\
            68 &10.6 &6.8 &8.1 &17.0 \\
            70  &8.9 &5.8 &6.7 &14.2 \\
            \hline
            \multicolumn{5}{c}{${}^{86}\text{Kr}+{}^{198}\text{Pt}$ system at $E_{\text{c.m.}}=324.2$~MeV} \\
            \hline
            60 &10.3 &6.7 &8.0 &16.7 \\
            80 &8.5 &5.6 &6.4 &13.7 \\
            100 &7.2 &4.8 &5.3 &11.5 \\
            120 &5.9 &4.0 &4.1 &9.2 \\
        \end{tabular}
    \end{ruledtabular}
    \label{tab_ca+pu_kr+pt_smf}
\end{table}

In the experiments of Refs.~\cite{kozulin2014c,kozulin2019,sen2022}, it
was observed that both quasi-fission and fusion-fission contribute to the
mass-symmetric region near $ ( \frac{A_{1}+A_{2}}{2}\pm 20 ) u \simeq (146\pm
20) $ for both systems. By using the quantal diffusion
approach (Eq.~\ref{eq_1}), we can determine the fragment mass distribution due
to the MNT mechanism (please see Eq.~(29) in the supplementary online material~\cite{supplemental} and
Ref.~\cite{risken1996}). Furthermore, to obtain information about the contribution
from fusion-fission-like events, we can utilize the statistical Monte Carlo code
GEMINI++~\cite{charity2008} for this region. The excitation energy of the
compound nuclei $^{284,292}\text{Fl}$ is estimated by,
$E_\mathrm{CN}^*=E_{\text{c.m.}}+Q_{gg}$, where $Q_{gg}$ stands for released
disintegration energy in fusion reaction. Combined with GEMINI++, total primary
fragment mass distribution takes the form
\begin{align}
\label{eq_4}
    Y(A)^{sum}=\left[\eta^{MNT}Y(A)^{MNT}+\eta^{FF}Y(A)^{FF} \right]\;,
\end{align}
where $Y(A)^{FF}$ stands for the probability of reaching fission-fragment with
mass number A, after the statistical de-excitation of superheavy flerovium
compound nucleus. The reliability of this method is discussed and verified in
the next section, Sec.~\ref{sec_4}. In GEMINI++ calculations, the number of
simulation times is set to $M_{\text{trial}}=100000$, which is sufficient to get
a statistical distribution for this region. In Eq.~(\ref{eq_4}), $\eta^{MNT}$ and
$\eta^{FF}$ stand for normalizing constants for distributions arising from the
MNT reaction and fission reaction, respectively. The value of $\eta^{MNT}$ is
determined by matching the peak values of experimental yield~\cite{kozulin2019}
at A$\approx$208 to give value $\eta^{MNT}=202$ for
${}^{48}\text{Ca}+{}^{244}\text{Pu}$ system and peak value of experimental
yield~\cite{sen2022} at A$\approx$200 to give value $\eta^{MNT}=225$ for
${}^{86}\text{Kr}+{}^{198}\text{Pt}$ system. Similarly, for fusion-fission
reactions, the value of $\eta^{FF}$ is determined by matching the experimental
yield at A$\approx$146 for both systems to give value $\eta^{FF}=17$ for
${}^{48}\text{Ca}+{}^{244}\text{Pu}$ system and $\eta^{FF}=4$ for
${}^{86}\text{Kr}+{}^{198}\text{Pt}$ system.

\section{Primary product mass distributions}\label{sec_4}
\begin{figure}[!htb]
  \centering
  \includegraphics[width=8.6cm]{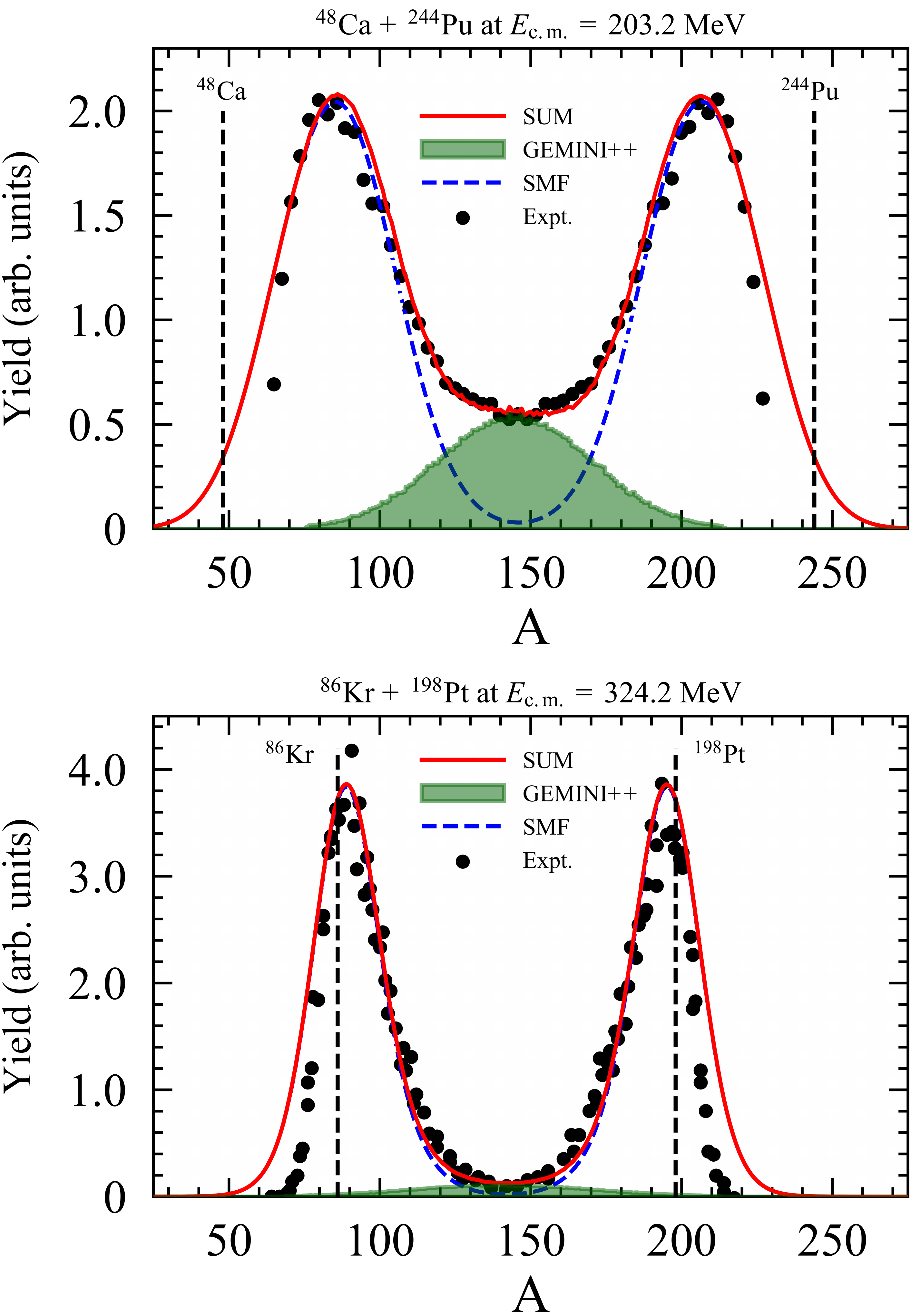}
  \caption{Primary fragment yield in the  ${}^{48}\text{Ca}+{}^{244}\text{Pu}$
system at $E_\text{{c.m.}}=203.2$~MeV and ${}^{86}\text{Kr}+{}^{198}\text{Pt}$
system at $E_\text{{c.m.}}=324.2$~MeV in tip orientation of target-like nucleus.
The experimental data obtained from Refs. ~\cite{kozulin2019,sen2022} are
indicated by the closed black circles. Dashed blue lines indicate the primary
product mass distributions calculated within the SMF framework. The
fusion–fission fragment distribution indicated by the green filled area was
calculated using the GEMINI++ code~\cite{charity2008}. The summation of yield
distribution calculated by the SMF approach and GEMINI++ are indicated by solid
red lines.}
  \label{fig_yield_combined}
\end{figure}

In Fig.~\ref{fig_yield_combined}, we share the calculated primary product
distributions for ${}^{48}\text{Ca}+{}^{244}\text{Pu}$ system at
$E_\text{{c.m.}}=203.2$~MeV and ${}^{86}\text{Kr}+{}^{198}\text{Pt}$ system at
$E_\text{{c.m.}}=324.2$~MeV in tip orientation of target-like nucleus and
compare the results with the experimental data
available~\cite{kozulin2019,sen2022}. Dashed blue lines indicate primary product
mass distribution calculated within the SMF framework, whereas filled green
areas represent the fusion–fission fragment distribution calculated using the
GEMINI++ code~\cite{charity2008}. The summation of yield distribution calculated
by Eq.~(\ref{eq_4}) is indicated by solid red lines. For comparison, the positions
of initial reaction partners are shown by vertical dashed black lines in
Fig.~\ref{fig_yield_combined}.

First, we can start by comparing the results obtained from the SMF approach for
both systems. In Fig.~\ref{fig_yield_combined}, we can see that the SMF approach
is well capable of explaining the mass distribution arising from the MNT
reactions near the peak values. In the vicinity of the peak position, around 40
nucleons are transferred for ${}^{48}\text{Ca}+{}^{244}\text{Pu}$ system, and
around 5 nucleons are transferred for the ${}^{86}\text{Kr}+{}^{198}\text{Pt}$
system. The SMF approach can clearly predict the peak position of the mass
distribution for both systems. Furthermore, in
${}^{48}\text{Ca}+{}^{244}\text{Pu}$ system, the SMF approach is able to explain
the transfer channels up to $\sim$55 nucleon transfer, up until A$\approx100$
for projectile-like fragments ($A\approx$ 190 for target-like fragments).
Similarly, in ${}^{86}\text{Kr}+{}^{198}\text{Pt}$ system, since the
contributions from fusion-fission are much lower, and MNT reactions are
dominant, the SMF approach agrees quite well with the experimental data.  We
have to mention that, since the primary mass probability $P_{\ell}(A)$ in
Eq.~(\ref{eq_2}) is normalized to unity, the value normalization constant $\eta$
also determines the integrated yield value under the mass distribution functions
in Fig.~\ref{fig_yield_combined}. In order to compare the contributions from
fusion-fission in these systems we can calculate the ratio of integrated yield
between the interval A$_{\text{CN}}\mp20$ to the total integrated yield
under the experimental values. In ${}^{48}\text{Ca}+{}^{244}\text{Pu}$
system, the interval for experimental data is taken as $65 \leq A \leq 225 $,
and the interval A$_{\text{CN}}\mp20$ region is taken as  $ 126 \leq A \leq
166$  to give the value $22.3/201.3 \approx \%11$. Similarly, in
${}^{86}\text{Kr}+{}^{198}\text{Pt}$ system, the interval for experimental data
is taken as $65 \leq A \leq 215 $, and the interval A$_{\text{CN}}\mp20$ region
is taken as  $ 122 \leq A \leq 162$ to give the value $6.5/225.4 \approx \%3$.
Both results are consistent with the experimental results reported in
Refs.~\cite{kozulin2019,sen2022}. This result clearly supports the idea that at
energies above the Bass barrier, the MNT reactions dominate fusion-fission
reactions with increasing Coulomb factor, $Z_PZ_T$.

\section{Conclusions}\label{sec_5}

In this work, we consider the charge and mass number as the macroscopic
variables and utilize the quantal diffusion approach based on the SMF theory to
calculate primary fragment mass distributions in  $^{48}$Ca+$^{244}$Pu reaction
at E$_\text{{c.m.}}= $ 203.2 MeV and $^{86}$Kr+$^{198}$Pt reaction at
E$_\text{{c.m.}} = $ 324.2 MeV. The quantal diffusion coefficients associated
with charge and mass variables are evaluated only in terms of the time-dependent
single-particle wave functions of TDHF theory. The description includes the full
collision geometry, quantal effects due to shell structure, and the Pauli
exclusion principle. We emphasize that the SMF theory does not involve any
adjustable parameters other than the standard parameters of energy density
functional employed in TDHF theory. The calculated primary product mass
distributions are compared with the experimental data for both systems. The
observed agreement between the experimental data and SMF results highlight the
effectiveness of the quantal diffusion approach based on the SMF approach.

\begin{acknowledgments}
S.A. gratefully acknowledges Middle East Technical University for the warm
hospitality extended to him during his visits. S.A. also gratefully acknowledges F. Ayik for continuous support and encouragement. This work is supported in part
by US DOE Grant Nos. DE-SC0015513 and DE-SC0013847.
This work is supported in part by TUBITAK Grant No. 122F150. The numerical
calculations reported in this paper were partially performed at TUBITAK ULAKBIM,
High Performance and Grid Computing Center (TRUBA resources).
\end{acknowledgments}

\bibliography{VU_bibtex_master}

\end{document}